# Onset of antiferromagnetism in UPt$_3$ via Th-substitution studied by muon spin spectroscopy

M. J. Graf,[1] A. de Visser,[2] C. P. Opeil,[1] J. C. Cooley,[3] J. L. Smith,[3] A. Amato,[4] C. Baines,[4] F. Gygax,[5] and A. Schenck[5]

[1]Department of Physics, Boston College, Chestnut Hill, MA 02467
[2]Van der Waals - Zeeman Institute, University of Amsterdam,
 1018 XE Amsterdam, The Netherlands
[3]Los Alamos National Laboratory, Los Alamos, NM 87545
[4]Laboratory for Muon-Spin Spectroscopy, Paul Scherrer Institute, CH-5232 Villigen, Switzerland
[5]Institute for Particle Physics of ETH Zürich, PSI CH-5232 Villigen, Switzerland

Muon spin spectroscopy has been used to study in detail the onset of large-moment antiferromagnetism (LMAF) in UPt$_3$ as induced by Th substitution. Zero-field experiments have been carried out on a series of polycrystalline U$_{1-x}$Th$_x$Pt$_3$ ($0 \leq x \leq 0.05$) samples in the temperature range 0.04 - 10 K. At low Th content ($x \leq 0.002$) magnetic ordering on the time scale of the μSR experiment ($10^{-8}$ s) is not detected. For $x = 0.005$ a weak magnetic signal appears below $T = 2$ K, while for $0.006 \leq x \leq 0.05$, spontaneous oscillations in the μSR spectra signal the presence of the LMAF phase. The data are well described by a two-component depolarization function, combining the contribution of a polycrystalline antiferromagnet and a Kubo-Lorentzian response. However, the transition into the antiferromagnetic phase is quite broad. For $x = 0.01$ and 0.02, a weak magnetic signal appears below about 7 K, which is well above the mean-field transition temperatures. The broadening may be a result of the effects of disorder on the time fluctuations associated with anomalous small-moment antiferromagnetism.

# I. INTRODUCTION

The compound UPt$_3$ is an exemplary unconventional superconductor.[1] Superconductivity emerges at $T_c \approx 0.5$ K from a strongly correlated heavy-fermion state that also exhibits an anomalous form of antiferromagnetism ("small-moment" antiferromagnetism, or SMAF) below ~ 6 K.[2] The superconducting state has a striking double transition[3] and multiple superconducting phases in applied magnetic fields.[4] Experiments have demonstrated that the magnetic SMAF couples to the superconducting state,[5,6] as expected for magnetically-mediated pairing. Knight shift[7] and substitutional studies[8,9] indicate that the superconducting state has odd parity. These experimental results are well accounted for by phenomenological Ginzburg-Landau theories, which make use of complex 1D or 2D order parameters. Multiple superconducting phases then arise from the lifting of intrinsic internal degeneracies of the order parameter by a symmetry-breaking field.[1] Pressure[10] and substitutional[11] studies indicate that the anomalous SMAF state is the symmetry-breaking field, although lattice distortions and defects[12-14] are possible alternatives.

While it is generally accepted that the pairing is due to magnetic correlations, the magnetic state from which superconductivity emerges is poorly understood. Neutron and magnetic x-ray scattering clearly signal a transition into the SMAF phase.[2,6] Additional Bragg scattering peaks appear at 6 K, which show that the small ordered magnetic moment is directed along the $a^*$ axis in the hexagonal plane (UPt$_3$ has a hexagonal MgCd$_3$-type of crystal structure). The magnetic unit cell consists of a doubling of the nuclear unit cell along the $a^*$ axis. The order parameter exhibits an anomalous quasi-linear temperature dependence. Bulk thermodynamic signatures of the SMAF phase are expected to be difficult to observe due to the small size of the ordered moment (0.02 $\mu_B$ as $T \rightarrow 0$ K), but sufficiently sensitive local probes such as NMR[15] and muon spin relaxation/rotation ($\mu$SR)[16,17] also do not signal a transition into the SMAF state. Thus it has been suggested that in UPt$_3$, SMAF is not a statically ordered phase. Rather, it fluctuates on time scales that are short compared to the typical NMR and $\mu$SR time scales (down to $10^{-8}$ s), yet are long compared to the nearly-instantaneous scales ($< 10^{-11}$ s) of neutron and x-ray scattering.[18] This picture qualitatively explains the discrepancy between the neutron measurements and $\mu$SR and NMR data. A similar picture has recently been proposed for the high-temperature superconductor YBa$_2$Cu$_3$O$_{6.5}$.[19]

Conventional antiferromagnetism with fairly large magnetic moments (large-moment antiferromagnetism, or LMAF) can be induced readily upon doping UPt$_3$. For example, upon substituting small amounts of Th for U pronounced phase transition anomalies are observed in the thermal and transport properties.[20,21] The λ-like anomaly in the specific heat and the chromium-type anomaly in the electrical resistivity give evidence for an antiferromagnetic phase transition of the spin-density wave type.[20] Antiferromagnetism in the U$_{1-x}$Th$_x$Pt$_3$ pseudobinaries has been detected in the concentration range $0.02 \leq x \leq 0.1$. The Néel temperature $T_N$ attains a maximum value of 6.5 K at about 5 at.% Th. Neutron-diffraction experiments[22] on single-crystalline U$_{0.95}$Th$_{0.05}$Pt$_3$ provide solid proof for antiferromagnetic ordering with an ordered moment of 0.65±0.1 μ$_B$/U-atom. Interestingly, the magnetic structures for the LMAF and SMAF are identical, indicating a close connection between the two types of magnetism.

The LMAF phase also appears in a very similar manner when Pt is replaced by small amounts of Pd[23] or Au.[24,25] This shows that the localization of the uranium moments is not governed by the unit cell volume of these pseudobinaries, as Th expands the lattice, while Pd and Au contract the lattice, but rather by the $c/a$ ratio,[24,26,27] which decreases for these dopants. This is corroborated by substitution studies[24,28] using small amounts of Ir (the $c/a$ ratio increases and no magnetic order is detected), as well as by pressure studies.[26,27] However, subsequent dopant studies[29] using Lu, Sc, Hf and Zr have cast some doubt on the role of the $c/a$ ratio as control parameter for the LMAF phase.

The goal of the present work is to investigate in detail the magnetic phase diagram of the U$_{1-x}$Th$_x$Pt$_3$ pseudobinaries. Our aim is to determine the LMAF phase boundary, and to study the possible interplay between the magnetic and superconducting phase. Our motivation stems from the close resemblance of the magnetic phase diagrams of U$_{1-x}$Th$_x$Pt$_3$ ($x \leq 0.15$)[21] and U(Pt$_{1-x}$Pd$_x$)$_3$.[26] The magnetic and superconducting properties of U(Pt$_{1-x}$Pd$_x$)$_3$ have been studied in great detail (for a review see Ref. 26). Recently, a series of neutron diffraction[30] and μSR[17,31] experiments, on polycrystalline as well as single-crystalline samples, led to the important finding that the critical concentration for the emergence of the LMAF phase, $x_{c,AFM}$, is approximately equal to the critical concentration for the suppression of superconductivity, $x_{c,SC}$. This was taken as evidence that static antiferromagnetism and odd-parity superconductivity are incompatible in this system, which in turn can be attributed to strong magnetic fluctuations in the vicinity of the quantum critical point at $x_{c,AFM} \approx x_{c,SC} \approx 0.006$.[31] The phase diagram for the U(Pt$_{1-x}$Pd$_x$)$_3$ compounds, delineating the superconducting, SMAF and LMAF phase, is shown in Fig.1. It will serve as a reference for the discussion of the results for Th-substituted compounds presented in this work.

In this paper we present a systematic µSR study of the magnetic phase diagram of U$_{1-x}$Th$_x$Pt$_3$ for several Th concentrations $0 \leq x \leq 0.05$. The motivation to use the µSR technique is given by the extreme sensitivity of the muon to weak magnetic signals. Additionally, the muon acts as a local probe and may be used to distinguish magnetically inequivalent sample regions and to determine the corresponding volume fractions. In a previous study[32], zero-field µSR experiments were carried out on polycrystalline U$_{1-x}$Th$_x$Pt$_3$ with $x = 0.01$ and 0.05. For $x = 0.05$ the LMAF phase appeared as a spontaneous oscillation in the µSR signal below $T_N = 6.5$ K. For $x = 0.01$, data were taken for $T > 4$ K only, and no magnetic phase transition was observed. Our new results extend over a large temperature range (0.04 - 10 K), as well as providing more detail on the Th concentration dependence.

## II. MUON DEPOLARIZATION FUNCTIONS

In this section we present the muon depolarization functions that are used to fit the µSR spectra obtained for the (U,Th)Pt$_3$ compounds. As will become clear in section IV, the magnetic properties of Th and Pd doped UPt$_3$ are very similar. Therefore, the fitting procedure described here relies to a great extent on a close parallel with the analysis of the µSR spectra of the U(Pt,Pd)$_3$ pseudobinaries.[17]

Zero-magnetic-field µSR is a local probe measurement of the magnetic field at the muon stopping site(s) in the sample.[33] If the implanted polarized muons are subject to magnetic interactions, their polarization becomes time dependent, $\boldsymbol{P}_\mu(t)$. By measuring the asymmetric distribution of positrons emitted when the muons decay as a function of time, the time evolution of $\boldsymbol{P}_\mu(t)$ can be deduced. The function $P_\mu(t)$ is defined as the projection of $\boldsymbol{P}_\mu(t)$ along the direction of the initial polarization: $P_\mu(t) = \boldsymbol{P}_\mu(t) \cdot \boldsymbol{P}_\mu(0) / P_\mu(0) = G(t) P_\mu(0)$. The depolarization function $G(t)$ reflects the normalized muon-spin autocorrelation function $G(t) = \langle \boldsymbol{S}(t) \cdot \boldsymbol{S}(0) \rangle / S(0)^2$, which depends on the average value, distribution, and time evolution of the internal fields, and it therefore contains all the physics of the magnetic interactions of the muon inside the sample.

In the LMAF phase, well below the ordering temperature ($T \ll T_N$), the depolarization is best described by a two-component function (compare Ref.17):

$$G(t) = A_{osc}\left[\frac{2}{3}\exp(-\lambda t)\cos(2\pi\nu t + \phi) + \frac{1}{3}\exp(-\lambda' t)\right] + A_{KL}G_{KL}(\lambda_{KL}t). \quad (1)$$

The first term on the right hand side is the time-dependent depolarization function for a polycrystalline antiferromagnet. This term consists of a 2/3 contribution from muons precessing with frequency $\nu$ in a static, non-zero, local magnetic field, and a 1/3 contribution from muons with spins effectively aligned parallel to the field ($\nu = 0$). The exponential decays $\lambda$ and $\lambda'$ reflect the distribution of local fields due to static variations and/or dynamical fluctuations. The second term is a Kubo-Lorentzian decay

$$G_{KL}(\lambda_{KL}t) = \frac{1}{3} + \frac{2}{3}(1 - \lambda_{KL}t)\exp(-\lambda_{KL}t). \quad (2)$$

In this case, the depolarization with characteristic rate $\lambda_{KL}$ is caused by an isotropic Lorentzian distribution of local fields with an average value of zero. For U(Pt,Pd)$_3$, the amplitudes of the two components of $G(t)$ were found to be equal, $A_{osc} \approx A_{KL}$, which indicates that the muon can stop at two distinct localization sites in the sample with equal probability.

For samples at $T > T_N$, the depolarization is found to result from the Gaussian distribution of static, randomly-oriented, magnetic fields due to $^{195}$Pt nuclei. As expected, the form of the depolarization function is given by the Kubo-Toyabe function

$$G_{KT}(\Delta_{KT}t) = A_{KT}\left[\frac{1}{3} + \frac{2}{3}(1 - \Delta_{KT}^2 t^2)\exp\left(-\frac{1}{2}\Delta_{KT}^2 t^2\right)\right], \quad (3)$$

with $\Delta_{KT} \approx 0.07$ μs$^{-1}$ (Ref.17). There is no zero-field μSR signature for the SMAF state,[16,17] so Eq. (3) works equally well in the paramagnetic phase as in the anomalous SMAF region.

Finally, in the vicinity of the Néel transition, the muon ensemble will be sensitive to sample domains possessing slightly different transition temperatures. The total depolarization will therefore be given in this temperature range by a sum of Eqs. (1) and (3). The breadth of the transition from LMAF to paramagnetism (or SMAF) is manifested in the amplitude of the magnetic component $A_M = A_{osc} + A_{KL}$ relative to the total signal, $A_{tot} = A_M + A_{NM}$. The non-magnetic component $A_{NM}$ is given by $A_{KT}$. The total signal amplitude is assumed to be constant, which is used as a fitting constraint.

### III. MATERIALS PREPARATION AND CHARACTERIZATION

Polycrystalline $U_{1-x}Th_xPt_3$ samples were prepared at Los Alamos National Laboratory with $x$ = 0.00, 0.002, 0.005, 0.006, 0.009, 0.01, 0.02 and 0.005. Two batches were prepared for $x$ = 0.02. As starting materials we used 99.99% pure U and Th, and 99.999% pure Pt. To ensure a homogeneous distribution of the relatively small amounts of Th in the sample, U was melted into the starting amount of Th in several increments. The stoichiometrically appropriate amount of Pt was then melted into the U-Th alloy in 5-6 increments. This sample was melted 8-10 times, broken into small pieces, and then re-melted another 8-10 times. The final product was annealed in a high-vacuum furnace (base pressure $6 \times 10^{-7}$ torr) at 850 °C for 5 days, followed by a slow cool-down (2 days). The annealed samples were spark cut into $6 \times 10 \times 1$ mm$^3$ plates that were polished for the µSR measurements. Additional smaller bars were cut for characterization via x-ray analysis, resistive measurements, etc.

X-ray studies on the samples show that there is less than 1% impurity phase in the samples. The lattice parameters of the undoped sample were $a = 0.5761(1)$ nm, $c = 0.4897(1)$ nm, and thus $c/a = 0.8501(3)$,[34] consistent with earlier results for UPt$_3$.[26] These values were found to be independent of Th concentration within the experimental error, as the resolution was not sufficient to track the anticipated very small changes in $c/a$ (less than 0.04 % for 5 at.% Th).[24]

The resistance was measured using a four-terminal ac-technique in a $^3$He refrigerator. The residual resistivities $\rho_0$ were determined by extrapolating the data between 1 K and 0.3 K (or $T_c$ for the samples with $x = 0$ and 0.002) to 0 K assuming a power-law temperature dependence.[11,35] $\rho_0$ varies smoothly across the series (see Fig. 2), indicating proper quality of the samples. Initially $\rho_0$ increases in a quasi-linear fashion, but for $x > 0.005$ the increase becomes super-linear. This behavior was also observed for U(Pt,Pd)$_3$,[26] and is believed to result from the opening of a gap at the Fermi surface upon formation of the LMAF spin-density wave state.

For samples with $x = 0.02$ and 0.05 we observe broad resistive anomalies near 5 K and 7 K, respectively, which signal the onset of the LMAF phase.[20] However, for samples with lower Th content ($x \leq 0.01$) no resistive anomalies associated with magnetic ordering were detected. The samples with $x = 0$ and 0.002 were found to have superconducting transition temperatures of 0.55 K and 0.43 K, respectively, consistent with earlier results.[36] For the samples with $x = 0.005$ and 0.006 no superconductivity was detected in the measured temperature range ($T > 0.3$ K).

## IV. μSR RESULTS FOR U$_{1-x}$Th$_x$Pt$_3$

Muon-spin spectroscopy measurements on U$_{1-x}$Th$_x$Pt$_3$ were carried out on the πM3 beamline at the Paul Scherrer Institute in Villigen, Switzerland. Measurements were conducted between 1.7 K and 10 K using the General Purpose Spectrometer, and between 0.04 K and 2.2 K using the Low Temperature Facility. Our primary results concern the muon-spin relaxation/rotation in zero applied magnetic field. The total depolarization amplitude (asymmetry) for each sample during a particular run was determined from measurements in a 100 G transverse magnetic field in the high-temperature paramagnetic state. In all the fitting procedures described below, the sum of the amplitudes of the fitting components is constrained to equal this total amplitude.

### A. $x \leq 0.002$

Zero-field experiments were carried out for the $x = 0$ sample in the temperature range 2-10 K. The muon depolarization is well-described by the Kubo-Toyabe function, with a temperature independent $\Delta_{KT} = 0.071(3)$ μs$^{-1}$. No sign of entry into the anomalous SMAF phase was observed near 6 K. This is consistent with earlier results on single-crystalline[16] and polycrystalline[17] un-doped UPt$_3$ samples. For x = 0.002, zero-field data were taken in the temperature range 0.05 K $\leq T \leq$ 0.9 K. The resulting fits with the Kubo-Toyabe equation yield $\Delta_{KT} = 0.103(5)$ μs$^{-1}$ independent of $T$. This behavior is consistent with the results for pure UPt$_3$. While $\Delta_{KT}$ is slightly larger than for our pure sample, an additional measurement made at 7.3 K yielded $\Delta_{KT} = 0.094(6)$

μs$^{-1}$. Therefore we rule out any weak increase of $\Delta_{KT}$ as temperature is lowered and conclude that the sample remains non-magnetic at all temperatures studied. These results are similar to observations made[17] on U(Pt$_{1-x}$Pd$_x$)$_3$ in the range $0 \leq x \leq 0.005$. We conclude that the local muon response is unaffected by entry into the SMAF phase, and is primarily due to $^{195}$Pt nuclear moments.

## B. $x \geq 0.006$

Zero-field data were obtained for samples with $x$ = 0.009, 0.01, 0.02 (two different samples) and 0.05 in the temperature range 1.8 - 10 K and for $x$ = 0.006 in the temperature range 0.04 - 3 K. For all samples we clearly observe the signature of static (on the typical muon timescale of $10^{-8}$ s) antiferromagnetism. In Fig. 3 we show the μSR spectra ($t < 0.6$ μs) for samples with $x$ = 0.01, 0.02 and 0.05 at the lowest temperature (1.8 K). For samples with $x \geq 0.009$ spontaneous oscillations are clearly observed and the oscillation frequency decreases with decreasing Th content. At low temperatures ($T \ll T_N$) the data are well described by the two-component depolarization function given in Eq. (1), i.e. the sum of depolarization due to a polycrystalline antiferromagnet and a Kubo-Lorentzian term. The solid lines in Fig. 3 illustrate the quality of the fit. The spectra obtained for the sample with x = 0.006 is satisfactorily described by Eq. (1) as well, but here the spontaneous oscillation is barely discernible, even at the lowest temperature ($T$ = 0.04 K). In Fig. 4 we summarize these results by plotting the relevant fit parameters obtained at $T \ll T_N$ as a function of Th concentration. The spontaneous oscillation frequency $\nu$, the decay rate $\lambda'$, and the Kubo-Lorentz $\lambda_{KL}$ decay time, all show a smooth concentration dependence, while the decay rate $\lambda$ increases dramatically with decreasing Th concentration. For comparison we have included in Fig. 4 the values reported for the LMAF phase in U(Pt,Pd)$_3$.[17] The similarity is evident.

The ratio of the amplitude of the oscillatory and the Kubo-Lorentz decay term, $A_{osc}/A_{KL}$, has been extracted from the fits using the constraint $A_{tot} = A_{osc}+A_{KL}$ and is presented in Fig. 5 as a function of Th concentration. Within the statistical error of the fits, we find that $A_{osc} = A_{KL}$ independent of thorium concentration, as was found for Pd-substituted samples.[17] This indicates that the muon localizes in the ordered phase at two distinct interstitial sites with equal probability. However, assigning high-symmetry interstitial sites to the two components in Eq. (1) is problematic. The first component with asymmetry $A_{osc}$ signals a stopping site with a fairly large local dipolar field, i.e. 0.19 - 0.33 T/$\mu_B$, as 5.4 MHz < $\nu$ < 9.5 MHz (Fig. 4). The second term with asymmetry $A_{KL}$ suggest the presence of a site where the local dipolar fields cancel or are at least smaller than ~ 0.01 T/$\mu_B$ as follows from the measured $\lambda_{KL}$. These measured values may be compared with the values calculated for the LMAF structure for several high symmetry interstitial sites presented in Ref. 37 (recall that the magnetic structures for Th and Pd doped UPt$_3$ are identical). Along (0,0,z) the calculated dipolar field is small ($B_{loc} = 0$ for $z = 0$), which indicates that muons localizing at this site give rise to the Kubo-Lorentzian term. Along (2/3,2/3,z) the calculated dipolar fields range between 0.26 and 0.67 T/$\mu_B$ (for an ordered moment of 1 $\mu_B$). This indicates that the oscillatory term should be attributed to this second axial symmetric site. However, this is not corroborated by transverse magnetic field μSR studies in the paramagnetic state of single-crystalline U(Pt$_{0.95}$Pd$_{0.05}$)$_3$.[38] The analysis of the Knight shift indicates a single muon localization site (0,0,0). This in turn has been taken as evidence that the sample is intrinsically inhomogeneous.[39] Moreover, comparison of the ordered moment determined by neutron diffraction and the μSR fit parameters for U(Pt,Pd)$_3$, show that $\lambda_{KL}$ scales with the ordered moment but $\nu$ does not.[17] Thus, the origin of the two-component response remains an unresolved problem, and the various parameters in Eqs. (1) and (2) are considered as phenomenological in nature.

We now discuss our results for the temperature dependence. In order to fit our time-dependent curves as $T$ increases towards $T_N$, we have incorporated the additional constraint that $A_{osc} = A_{KL}$ (in addition to fixing the total amplitude), consistent with our experimental results at $T \ll T_N$. Also, we found that as $T$ increases towards $T_N$ use of Eq. (1) is not sufficient, and we have used the sum of Eqs. (1) and (3). The Kubo-Toyabe relaxation rate $\Delta_{KT}$ was extracted from the high-temperature data ($T > T_N$) and used as a constant in the fitting procedure. This value falls in the 0.05 - 0.08 μs$^{-1}$ range for all the samples studied, apart from the previously mentioned $x = 0.002$.

Analysis of the $x = 0.05$ data is straightforward. The two-component fit is found to work quite well except in a narrow region within about 1 K of the Néel temperature. Just like for U(Pt,Pd)$_3$ we find that the temperature dependence of the parameters $\nu$ and $\lambda_{KL}$ is mean-field-like and can be fit to the form[17]

$$P = P(0)\left[1-\left(\frac{T}{T_N}\right)^{\alpha}\right]^{\beta}, \qquad (4)$$

where $T_N$ is the Néel temperature, and $P$ is either $\nu$ or $\lambda_{KL}$. The resulting temperature dependence of $\nu$ and $\lambda_{KL}$ is plotted for our $x = 0.05$ sample in Fig. 6. Using Eq. (4) to fit the data yields $T_N = 7.02(2)$ K for both curves; $\nu(0) = 9.8(5)$ MHz, $\alpha = 2.0(6)$, and $\beta = 0.42(5)$ from $\nu(T)$, and $\lambda_{KL}(0) = 6.7(3)$ µs$^{-1}$, $\alpha = 3(1)$, and $\beta = 0.35(7)$ from $\lambda_{KL}(T)$. These values for the exponents compare favorably with those extracted in Ref. 17 for U(Pt$_{0.95}$Pd$_{0.05}$)$_3$: $\alpha = 2.1(3)$ and $\beta = 0.39(2)$ from $\nu(T)$, and $\alpha = 2.0(5)$ and $\beta = 0.36(6)$ from $\lambda_{KL}(T)$. As pointed out in that work, the values are consistent with the theoretical prediction of $\beta = 0.38$ derived for the 3D Heisenberg model[40] and $\alpha = 2$ calculated for a cubic antiferromagnet.[41] We also note that our values for $\nu(T)$ are in excellent agreement with the results for U$_{0.95}$Th$_{0.05}$Pt$_3$ reported in Ref. 32. The mean-field $T_N$ value of 7.0 K is in good agreement with thermal and transport measurements.[20,21]

A similar analysis for the temperature evolution of the muon response is problematic for samples with $0.006 \leq x \leq 0.02$. This is mainly due to the strong damping of the oscillatory term, as is illustrated by the large values of the decay $\lambda$ for $x \leq 0.01$ shown in Fig. 4b. Upon increasing the temperature, an accurate determination of the fitting parameters using Eq. (1) becomes more and more difficult. Moreover, as will be discussed in the next paragraph, the phase transition regions are quite broad and in the case of $x = 0.01$ and $0.02$ extend up to 7 K. Therefore, we have used the sum of Eqs. (1) and (3) in a broad temperature window up to ~ 7 K. For sample $x = 0.02$, the results of $\nu(T)$ and $\lambda_{KL}(T)$, are shown in Fig. 6. Notice that a few data points for $T > 5.5$ K have been omitted because of their large error bars. The fit parameters obtained for two different samples are in excellent agreement. Only $\nu(T)$ obeys mean-field behavior as expressed in Eq. (4), albeit in a limited temperature interval. When fitting $\nu(T)$ for $x = 0.02$ and $T < 4$ K to Eq. (4) using the model values $\alpha = 2$ and $\beta = 0.38$, we find $T_N = 5.05(5)$ K. The fit is shown in Fig. 6. This mean-field value of $T_N$ is in fair agreement with specific heat data.[20,21] While $\lambda_{KL}(T)$ decreases slightly as $T$ approaches $T_N$, it could not be fit using Eq. (4). We believe this is related to the magnetic inhomogeneity present in the system, as discussed in the next paragraph. A similar procedure for the $x = 0.01$ sample yields $T_N = 3.50(5)$ K. In Ref. 32, U$_{0.99}$Th$_{0.01}$Pt$_3$ was also studied via µSR, and no evidence of magnetic behavior was reported. However, it should be noted that the data for that sample were limited and extended only down to approximately 4 K.

The most striking aspect of the data is that our depolarization curves unambiguously show that magnetism is present in the system at temperatures well above the mean-field values for $T_N$. This is shown in Fig. 7, where the time-dependent polarization for $x = 0.02$ is plotted over short times at temperatures of 5.3 K, 5.8 K, and 7.4 K, all of which are greater than the mean-field $T_N$ of 5.05 K. While extraction of the exact parameters can be difficult when the condition $T \ll T_N$ is no longer met, we can readily characterize the transition width by calculating the magnetic fraction of the total amplitude, i.e., the fraction of the total signal due to the depolarization described by Eq. (1): $A_M/A_{tot} = (A_{osc}+A_{KL})/A_{tot}$. We have plotted $A_M/A_{tot}$ in Fig. 8 for samples with $x = 0.01$, 0.02, and 0.05. For comparison, we have also plotted the same quantity as extracted from the data of Ref. 17 for Pd-substituted samples. All the Pd-substituted samples have narrow transition widths, as does the Th-substituted sample with $x = 0.05$. However, the transitions for the Th-substituted samples with $x = 0.01$ and 0.02 are quite broad and have a magnetic component up to $T = 7$ K, which is the transition temperature for $x = 0.05$. It is clear that some form of magnetic inhomogeneity is present. The second sample with $x = 0.02$, independently fabricated at a later date, yields essentially identical data, confirming this behavior.

The validity of the assumption that $A_{osc} = A_{KL}$, and of the actual form of the fitting functions themselves, is debatable in the broad transition regions. However, the calculation of the fractional magnetic signal is insensitive to the functional form of the magnetic component. We have used a variety of different fitting functions to describe the magnetic contribution in the vicinity of $T_N$, and always reproduce the qualitative features shown in Figure 8.

### C. $x = 0.005$

We have also studied a sample with $x = 0.005$, which showed strong depolarization at low temperatures, but no oscillations were observed due to the heavy damping, as $\lambda$ increases dramatically with decreasing Th concentration (see Fig. 4b). In order to follow the temperature dependence of the magnetism, Eq. (1) was modified by replacing the two oscillatory terms with two exponential decays (i.e., setting the frequency equal to zero). An analysis similar to the one presented in the preceding paragraph shows that once again the transition from magnetic to non-magnetic behavior is quite broad, as shown in Fig. 9. It is important to note that for this sample the fractional magnetic signal $A_M/A_{tot}$ appears to reach a maximum value of about 0.6 at 0.085 K, suggesting that the sample never has a fully formed LMAF state.

## V. DISCUSSION

We conclude from the low-temperature ($T \ll T_N$) results that the two-component response function, Eq. (1), yields a proper description of the depolarization within the LMAF phase of our Th doped samples. The same depolarization function, with comparable values for the fit parameters and relative amplitudes of the different components, describes the LMAF phase in U(Pt,Pd)$_3$. Thus Eq. (1) appears to be a general characteristic of the LMAF phase, independent of whether the substitution is on the uranium or platinum sub-lattice. Note that Th substitution is far more effective in inducing LMAF than Pd substitution, since in our notation equal values of $x$ translate into a number of Pd impurities that is three times the number of Th impurities. These observations are consistent with the assumption that the $c/a$ ratio is the controlling factor in the onset of LMAF magnetism, since comparable values of $x$ in the case of Th and Pd doping yield comparable changes in $c/a$.[26]

The amplitudes $A_{osc}$ and $A_{KL}$ are found to be independent of concentration and impurity type. This important observation rules out an interpretation of the two-components being due to two different stopping sites where muons experience either a local field due to sites with all U (or Pt) nearest neighbors or sites where one or more of the U (Pt) nearest neighbors have been replaced by Th (Pd).

The most striking difference between our data for Th-substituted samples and similar data for Pd-substituted samples is the very broad magnetic transition region, as shown in Fig. 8. X-ray diffraction analysis shows that there is less than 1% impurity phase in the samples, and it is known from earlier work that U$_{1-x}$Th$_x$Pt$_3$ remains single phase up to about $x = 0.25$.[21] Strong variations in the local Th concentration across the sample volume could conceivably produce the broad transitions, since the Néel temperature depends on $x$. Such an effect would be consistent with the sharp transition observed for $x = 0.05$ because the $T_N(x)$ curve is expected to be fairly flat near $x = 0.05$ (see Fig. 1), and so a given distribution of concentrations would produce a less-broad transition region. However, the transition regions for the $x = 0.01$ and 0.02 samples extend up to 7 K. For the $x = 0.01$ sample, we estimate that about 10 % of the sample volume would need to have a local Th concentration well above $x = 0.02$ in order to reproduce the broadening shown in Fig. 9. This would severely deplete other regions, yet the data shown in Fig. 8 show that the sample has a negligible volume fraction with Néel temperature less than 2 K. Moreover, such dramatic inhomogeneity is inconsistent with smooth variation of the residual resistivity shown in Fig. 2. Thus chemical inhomogeneity seems unlikely to be the cause of the observed broadening.

An alternative explanation for the broadening of magnetic transition involves the effect of Th disorder on the anomalous SMAF phase. If the SMAF phase is indeed a time-fluctuating version of the LMAF phase, then Th impurities may serve to slow down the fluctuations. When the fluctuation timescale becomes comparable to, or longer than, the typical muon spectroscopy timescales one would expect to observe a magnetic signal. If so, the muon measurements would signal magnetic behavior near the onset temperature of the SMAF phase. No neutron diffraction studies have been carried out on the concentration dependence of either the SMAF or LMAF phases in $U_{1-x}Th_xPt_3$. However, neutron diffraction results for $U(Pt,Pd)_3$[30] show that SMAF is robust upon alloying: the transition into the SMAF state remains fixed at 6 K with increasing $x$ (up to $x = 0.01$), although the ordered moment increases with $x$. By analogy we expect that the SMAF phase line for (U,Th)Pt$_3$ will be essentially independent of Th concentration and fixed at a value of approximately 6 – 7 K. Thus it would be expected that for strong disorder, one could observe magnetism beginning at about 7 K, as observed in this work. This scenario could also explain the discrepancy between recent μSR studies on UPt$_3$[16,17] and the much earlier work by Heffner and co-workers.[32] In Ref. 32 a small increase in the zero-magnetic field depolarization rate was observed at the SMAF transition temperature of 6 K. However, later work on single crystals[16] and polycrystals[17] of high quality gave no evidence of the transition. It is possible that the sample quality for the work described in Ref. 32 was such that impurities played a role in slowing down the SMAF fluctuations, rendering the transition observable. At present we have no explanation as to why Th is apparently much more effective than Pd in slowing down the SMAF fluctuations.

It is also difficult to reconcile the growing body of data probing the SMAF-to-LMAF transition with the phase diagram shown in Fig. 1. Our μSR results imply that the transition is not abrupt, but results from a slowing down of the SMAF oscillations. Moreover, recent studies utilizing cantilever magnetometry[42] with a characteristic frequency of 1 kHz do not show a magnetic transition in U(Pt$_{0.99}$Pd$_{0.01}$)$_3$ single crystals, despite what appears to be sufficient sensitivity and even though µSR measurements[17] clearly indicate $T_N$ = 1.8 K. These results imply that the SMAF-to-LMAF transition is not a true phase transition but rather a type of crossover behavior. The details of a phase diagram such as that shown in Fig. 1 will depend on the characteristic timescale of the measuring probe, at least in some critical crossover region.

It is difficult to test for the existence of a superconducting/antiferromagnetic mutual quantum critical point, as found for U(Pt,Pd)$_3$.[31] Apart from the $x$ = 0.05 sample, an unambiguous determination of the Néel temperature is not possible from our data. For the $x$ = 0.005 sample, it is expected that $T_c$ = 0.2 K,[36] while we clearly observe magnetic behavior below about 2 K. This would seem to rule out the possibility of $x_{c,SC} \approx x_{c,AFM}$. However, for $x$ = 0.005 the magnetic signal is not developed in the whole sample volume as $A_M/A_{tot}$ approaches 0.6 as $T \to 0$ K (Fig. 9). This magnetic volume fraction is attributed to the LMAF phase, as it is entirely due to the two-component depolarization response, albeit with $v$ = 0. Therefore, superconductivity may occupy the remaining ~ 40% of the sample volume and thus still compete with the LMAF state. More detailed studies for samples in the vicinity of $x$ = 0.005 are required to clarify the relationship between superconductivity and LMAF.

The magnetic inhomogeneity in (U,Th)Pt$_3$, as evidenced by the two-component muon response function and the broad SMAF-to-LMAF transition, is particularly interesting since it was recently discovered that URu$_2$Si$_2$, another U-based small-moment heavy fermion system is magnetically inhomogeneous. NMR[43] and µSR[44] measurements under applied pressure showed that the small moment is caused by a small fraction of the sample volume having a relatively large local ordered moment, while the majority of the sample is paramagnetic. Neutron scattering[45] yields an ordered moment that is averaged over the entire volume of the sample. While there are significant differences between the two systems, it is clearly of interest to further probe the possibility of magnetic inhomogeneity in (U,Th)Pt$_3$ and U(Pt,Pd)$_3$ with an eye to similarities with the URu$_2$Si$_2$ system.

## VI. SUMMARY

In summary, we have used muon spin spectroscopy to study the onset of the large-moment antiferromagnetic phase (LMAF) in UPt$_3$ as induced by Th-substitution. At low Th content ($x \leq$ 0.002) magnetic ordering on the time scale of the μSR experiment ($10^{-8}$ s) is not detected, as is the case for pure UPt$_3$. For $0.006 \leq x \leq 0.05$, spontaneous oscillations in the μSR spectra signal the presence of the LMAF phase. The data are well described by the sum of two depolarization functions, namely a contribution from a polycrystalline antiferromagnet and a Kubo-Lorentzian response. This two-component depolarization function was previously used to describe the muon response in the LMAF phase of pseudobinary U(Pt,Pd)$_3$. However, the transition into the antiferromagnetic phase as temperature is lowered is much broader for Th substitution than for Pd substitution. The broad transition makes it difficult to detail the competition between superconductivity and LMAF in (U,Th)Pt$_3$, however it may provide an important clue as regards the nature of the SMAF phase. For $x = 0.01$ and 0.02 the magnetic signal extends up to ~ 7 K, which suggests that the broadening may be a result of the effects of disorder on the time fluctuations associated with the anomalous antiferromagnetic state (SMAF). These results imply that SMAF-to-LMAF is not a true phase transition but rather a crossover behavior. We are currently conducting detailed materials analysis and thermodynamic studies to test for this possibility.

## ACKNOWLEDGMENTS

We thank Eric J. Peterson of Los Alamos National Laboratory for carrying out x-ray diffraction measurements on the samples used in this study. μSR experiments were performed at the Swiss Muon Source, Paul Scherrer Institute, Villigen. This work was supported by the Petroleum Research Fund of the American Chemical Society, DOE grant DEF 60202ER63404, and the FERLIN program of the European Science Foundation. Work in Los Alamos was performed under the auspices of the U.S.D.O.E.

**FIGURE CAPTIONS**

Fig. 1  The superconducting and magnetic phase diagram of $U(Pt_{1-x}Pd_x)_3$, adapted from Ref. 31. SC = superconductivity, SMAF = small-moment antiferromagnetism, LMAF = large-moment antiferromagnetism. The phase boundary for the LMAF phase has been observed by thermal and transport measurements, as well as by neutron diffraction and µSR. The phase line for SMAF is observed by neutron diffraction only.

Fig. 2  Variation of the residual resistivity of $U_{1-x}Th_xPt_3$. The two values for $x = 0.02$ are for independently fabricated samples. The solid line is a guide to the eye.

Fig. 3  The short-time depolarization as a function of time for several $U_{1-x}Th_xPt_3$ samples taken at low-temperature ($T = 1.8$ K). Curves are displaced along the vertical axis for sake of clarity. Solid lines represent fits to the data using Eq. (1).

Fig. 4  Fitting parameters of the two-component depolarization function Eq. (1) as function of impurity content $x$ in $UPt_3$. Filled symbols are for $U_{1-x}Th_xPt_3$, open symbols for $U(Pt_{1-x}Pd_x)_3$.[17] All values are determined at $T = 1.8$ K, except the values for $x = 0.006$, which are evaluated at 0.1 K.

Fig. 5  Variation of the ratio of the asymmetries of the oscillatory component, $A_{osc}$, and the Kubo-Lorentzian component, $A_{KL}$, with Th concentration (see Eq. (1)).

Fig. 6  Temperature dependence of the (a) spontaneous oscillation frequency, and (b) Kubo-Lorentz damping factor, for $U_{0.95}Th_{0.05}Pt_3$ and $U_{0.98}Th_{0.02}Pt_3$. The solid lines are the mean-field fits, as described in the text. The squares and triangles for the $x = 0.02$ denote results for two independently fabricated samples.

Fig. 7  Temperature evolution of the depolarization function at short times for $U_{0.98}Th_{0.02}Pt_3$. Note that all temperatures are above the mean-field Néel temperature of 5.05 K.

Fig. 8  Transition widths as illustrated by the temperature-dependent fractional amplitude associated with magnetism. Circles are for $U_{1-x}Th_xPt_3$ while triangles are for $U(Pt_{1-x}Pd_x)_3$ with equivalent *x* values, taken from the work of Ref.17. Solid and dashed lines are guides to the eye for the Th-substituted and Pd-substituted data, respectively.

Fig. 9  Temperature dependence of the fraction of the total amplitude of the depolarization function associated with magnetic behavior for $U_{0.995}Th_{0.005}Pt_3$.

Figure 1

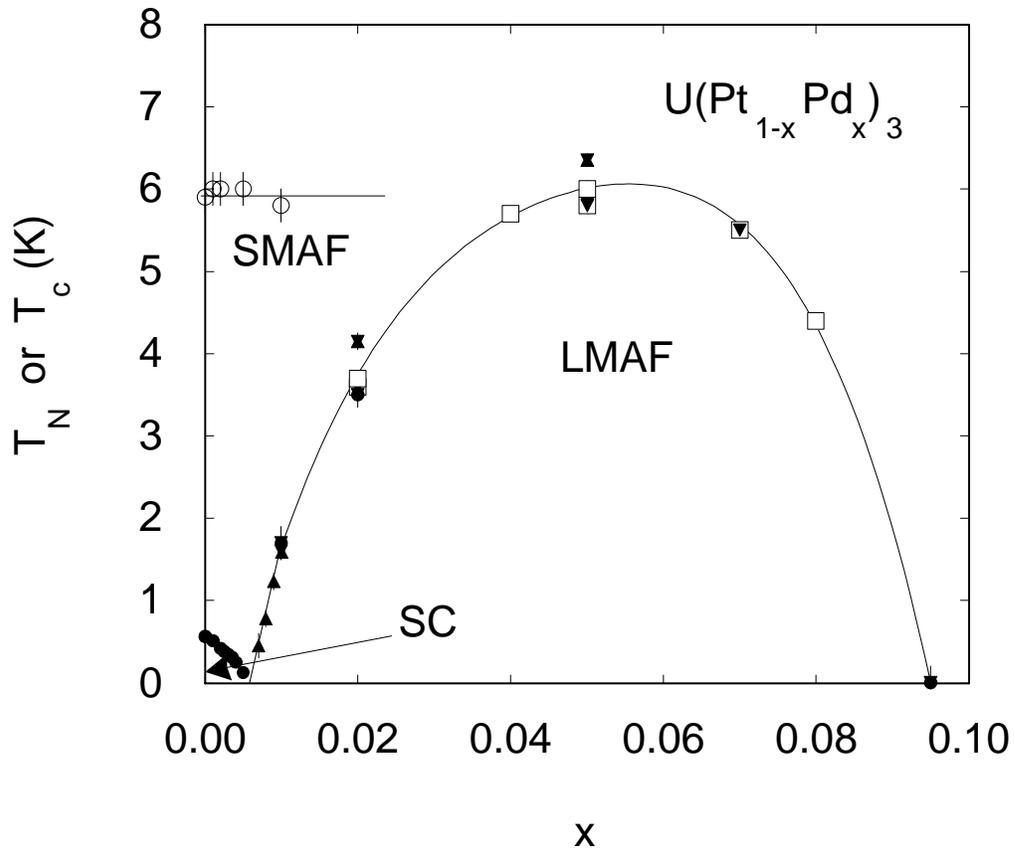

Figure 2

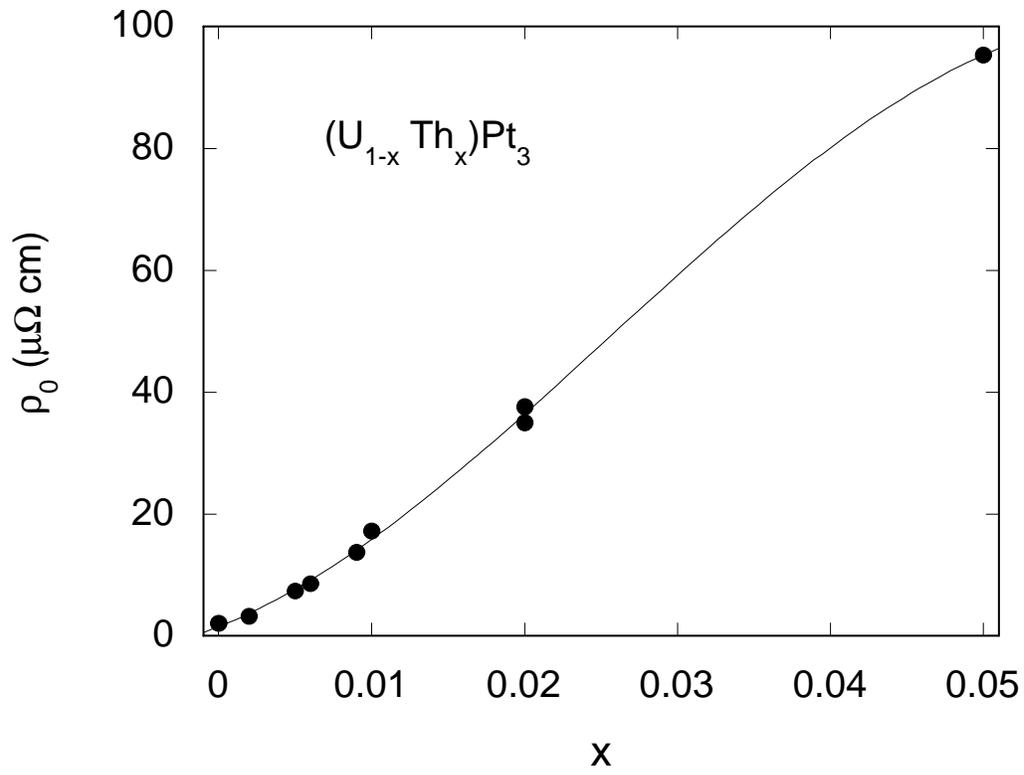

Figure 3

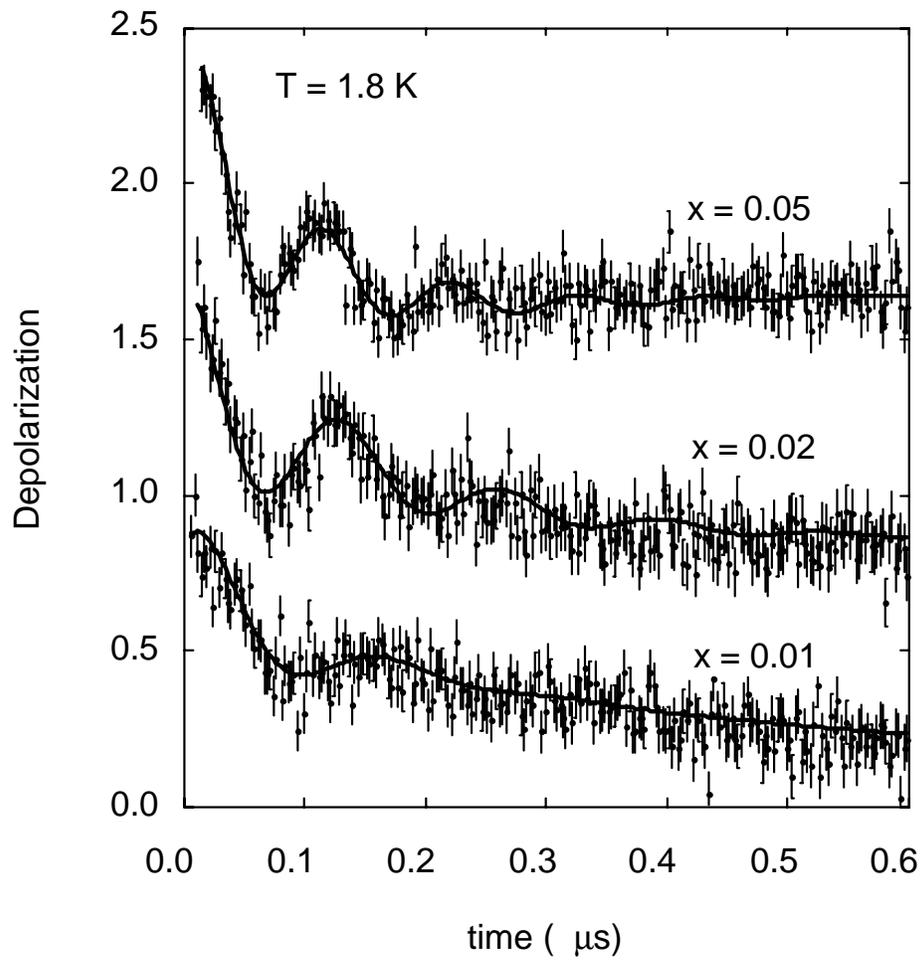

Figure 4

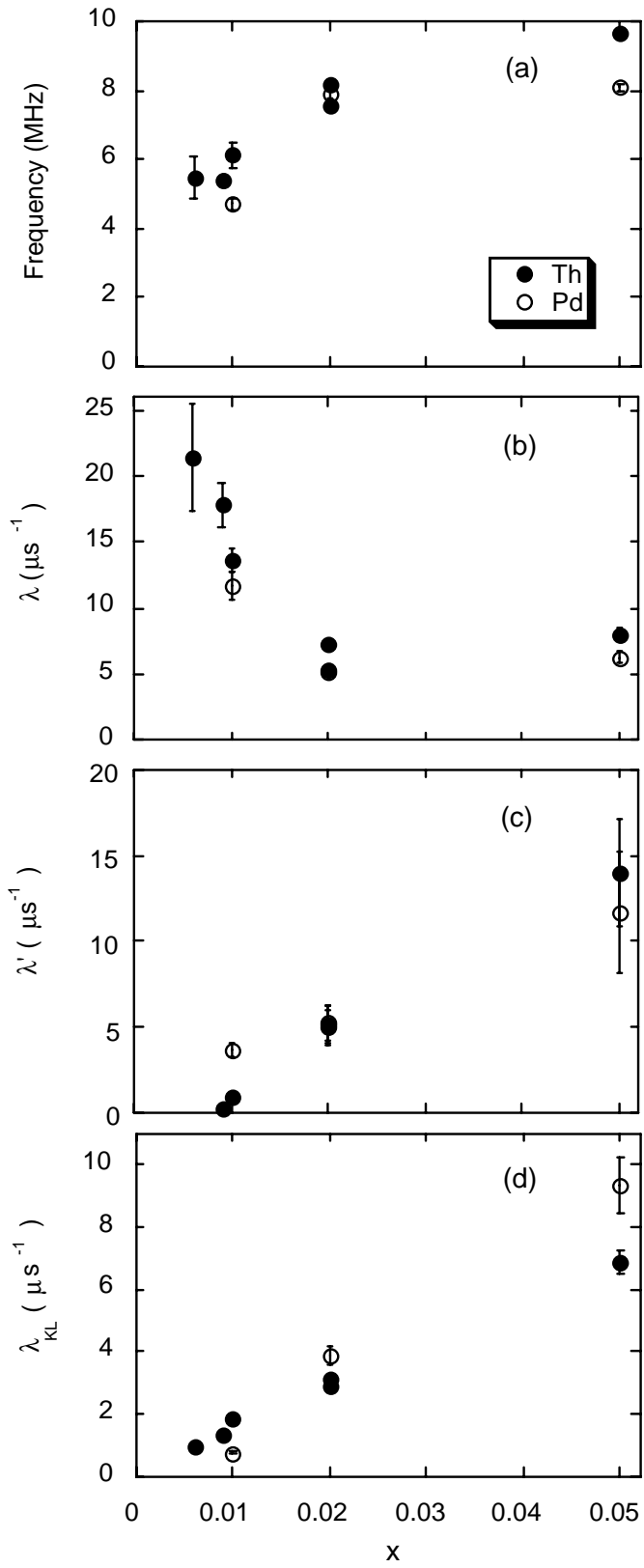

Figure 5

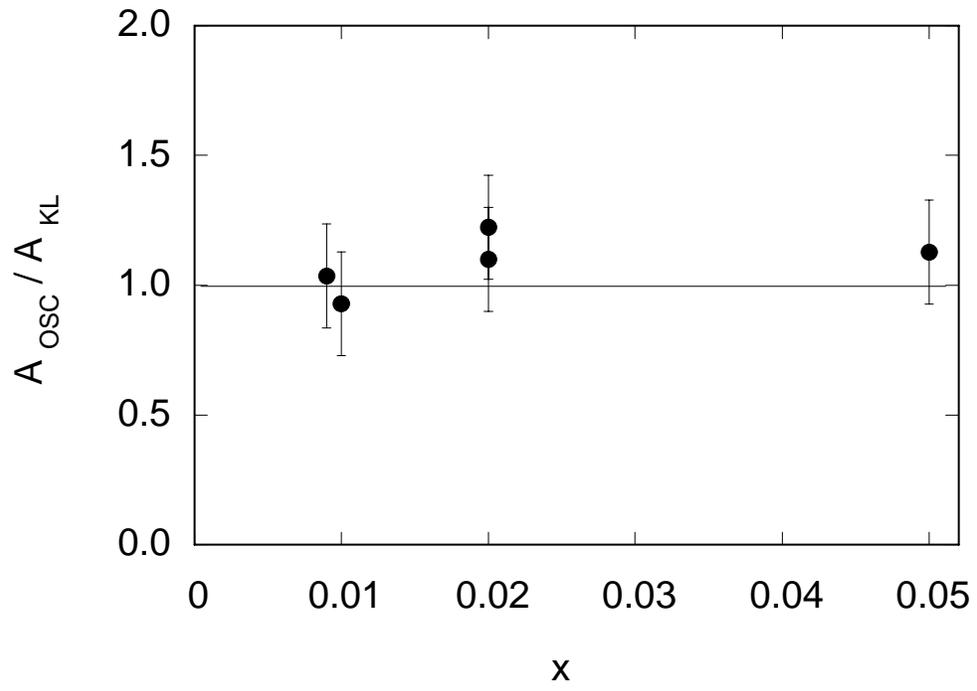

Figure 6

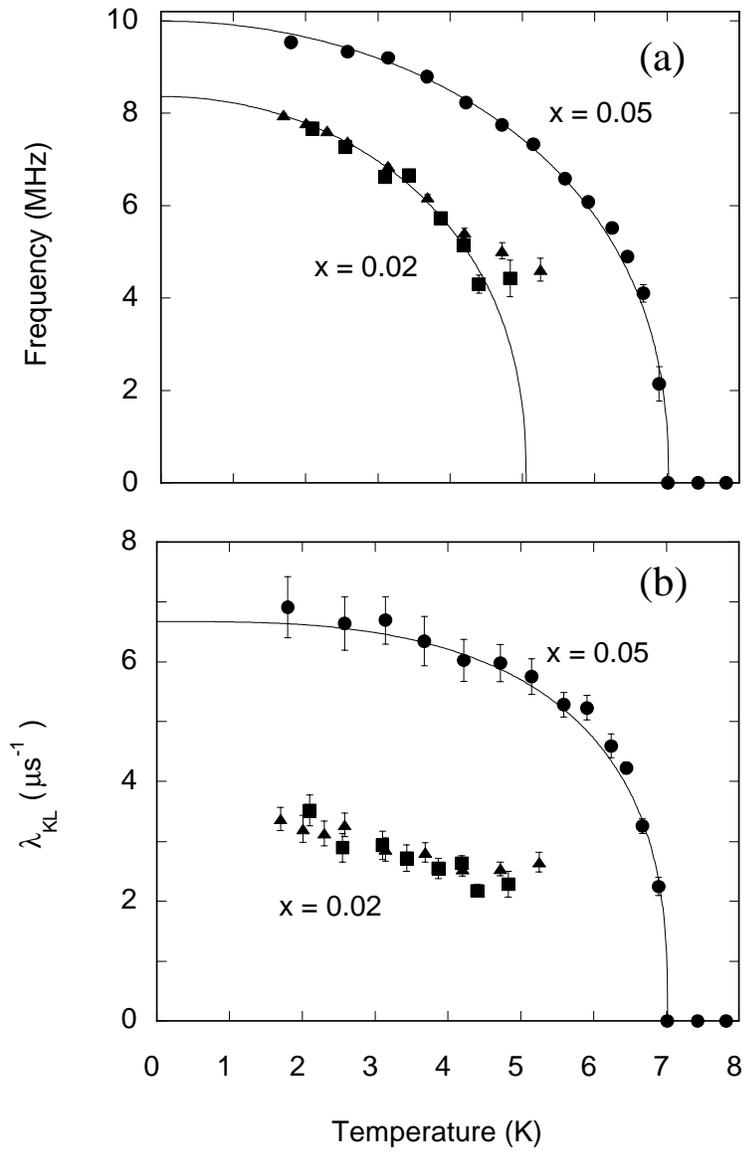

Figure 7

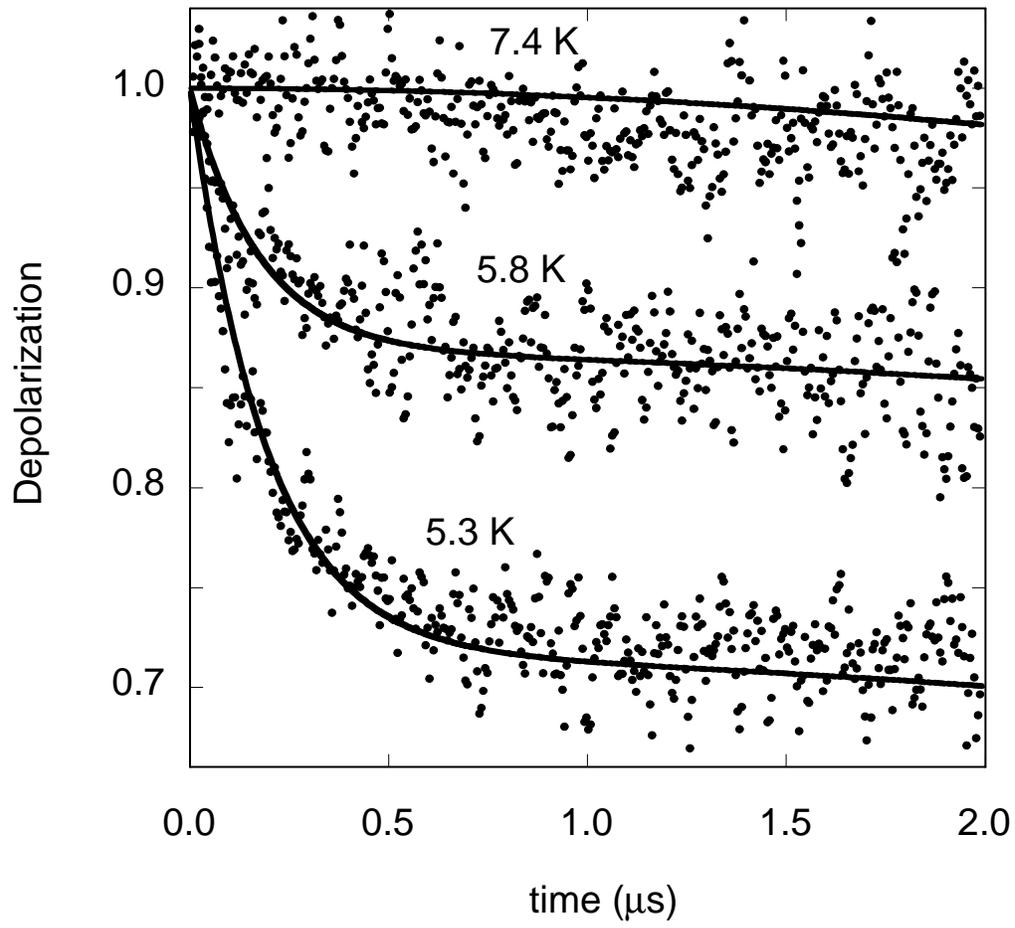

Figure 8

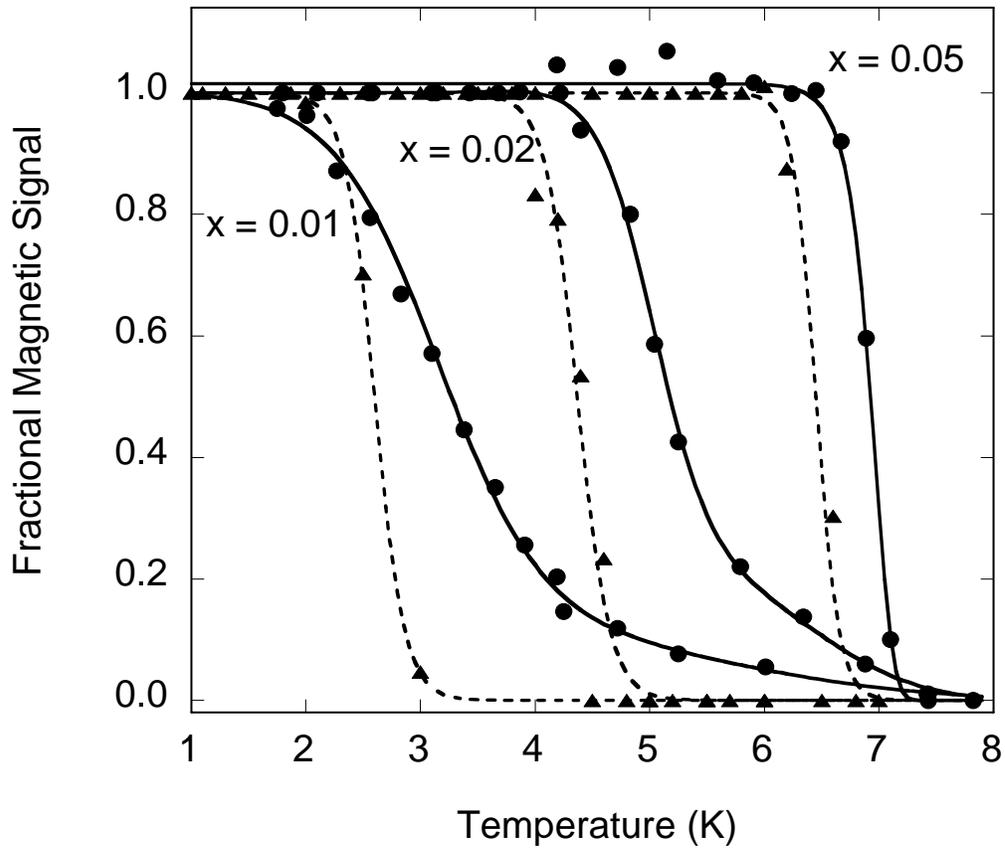

Figure 9

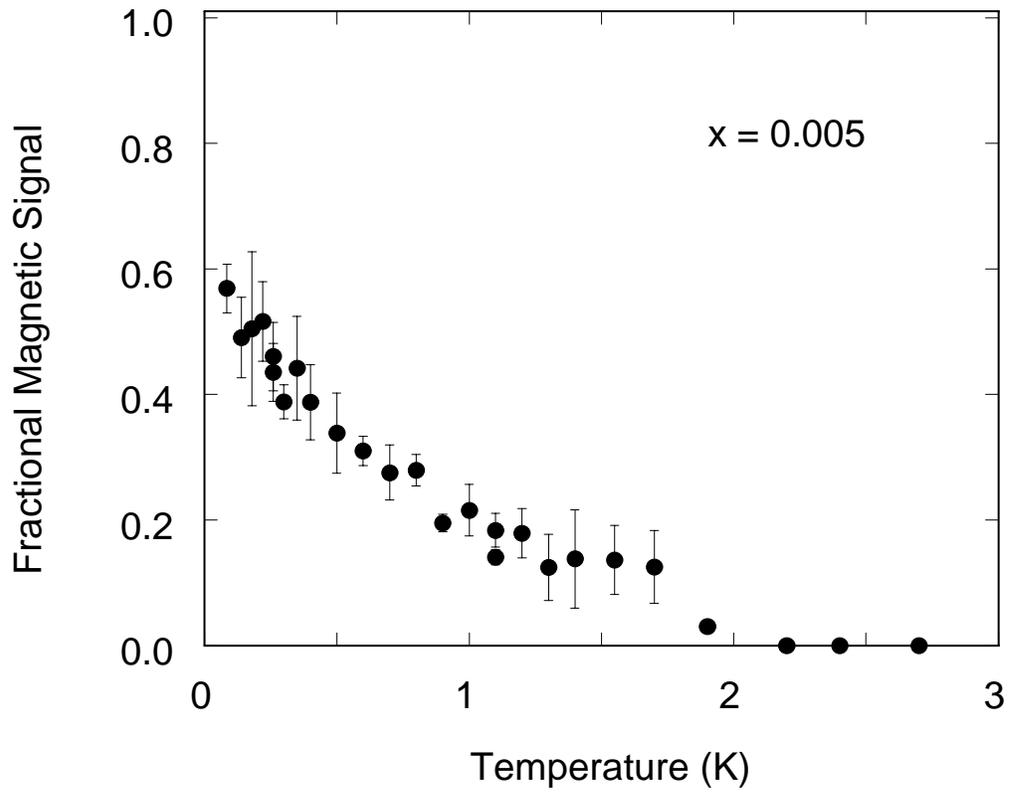